\documentclass[12pt, a4paper]{article}
\usepackage{graphicx}
\begin{document}

\title{The history of the cosmological constant problem
\footnote{Invited talk at the XVIIIth IAP Colloquium:
Observational and theoretical results on the accelerating
universe, July 1-5 2002, Paris, France }}

\author{Norbert Straumann\\
        Institute for Theoretical Physics University of Zurich,\\
        CH--8057 Zurich, Switzerland}
\date{\today}

\maketitle

\begin{abstract}
The interesting early history of the cosmological term is
reviewed, beginning with its introduction by Einstein in 1917 and
ending with two papers of Zel'dovich, shortly before the advent of
spontaneously broken gauge theories. Beside classical aspects, I
shall also mention some unpublished early remarks by Pauli on
possible contributions of vacuum energies in quantum field theory.
\end{abstract}

\section{Introduction}

One of the contributions in the famous volume {\it Albert
Einstein: Philosopher--Scientist} \cite{1} is an article by George
E. Lema\^{\i}tre entitled {\it ``The Cosmological Constant.''} In
the introduction he says: {\it ``The history of science provides
many instances of discoveries which have been made for reasons
which are no longer considered satisfactory. It may be that the
discovery of the cosmological constant is such a case.''} When the
book appeared in 1949 -- at the occasion of Einstein's seventieth
birthday -- Lema\^{\i}tre could not be fully aware of how right he
was, how profound the cosmological constant problem really is,
especially since he was not a quantum physicist.

During this week we shall hear why we are indeed confronted with a
deep mystery and the current evidence for the unexpected finding
that the recent ( $z<1$) Universe is dominated by an exotic
homogeneous energy density with {\it negative} pressure will be
summarized by various speakers. The simplest candidate for this
exotic energy density is a cosmological term in Einstein's field
equations, a possibility that has been considered during all the
history of relativistic cosmology. The organizers of this
interesting meeting asked me to review the main aspects of the
history of the $\Lambda$-term, from its introduction in 1917 up to
the point when it became widely clear that we are facing a deep
mystery.

\section{Einstein's original motivation of the $\Lambda$-term}

The cosmological term was introduced by Einstein when he applied
general relativity for the first time to cosmology. In his paper
of 1917 \cite{2} he found the first cosmological solution of a
consistent theory of gravity. In spite of its drawbacks this bold
step can be regarded as the beginning of modern cosmology. It is
still interesting to read this paper about which Einstein says:
{\it ``I shall conduct the reader over the road that I have myself
travelled, rather a rough and winding road, because otherwise I
cannot hope that he will take much interest in the result at the
end of the journey.''} In a letter to P. Ehrenfest on 4 February
1917 Einstein wrote about his attempt: {\it ``I have again
perpetrated something relating to the theory of gravitation that
might endanger me of being committed to a madhouse. (Ich habe
wieder etwas verbrochen in der Gravitationstheorie, was mich ein
wenig in Gefahr bringt, in ein Tollhaus interniert zu werden.)''}
\cite{3}.

In his attempt Einstein assumed -- and this was completely novel
-- that space is globally {\it closed}, because he then believed
that this was the only way to satisfy Mach's principle, in the
sense that the metric field should be determined uniquely by the
energy-momentum tensor. In these years and for quite some time
Mach's ideas on the origin of inertia played an important role in
Einstein's thinking. This may even be the primary reason that he
turned so soon after the completion of general relativity to
cosmology. Einstein was, in particular, convinced that isolated
masses cannot impose a structure on space at infinity. It is along
these lines that he postulated a universe that is spatially finite
and closed, a universe in which no boundary conditions are needed.
Einstein was actually thinking about the problem regarding the
choice of boundary conditions at infinity already in spring 1916.
In a letter to Michele Besso from 14 May 1916 he also mentions the
possibility of the world being finite. A few month later he
expanded on this in letters to Willem de Sitter.

 In addition, Einstein assumed that the Universe was {\it static}. This was not
unreasonable at the time, because the relative velocities of the
stars as observed were small. (Recall that astronomers only
learned later that spiral nebulae are independent star systems
outside the Milky Way. This was definitely established when in
1924 Hubble found that there were Cepheid variables in Andromeda
and also in other galaxies. Five years later he announced the
recession of galaxies.)

These two assumptions were, however, not compatible with
Einstein's original field equations. For this reason, Einstein
added the famous $\Lambda$-term, which is compatible with the
principles of general relativity, in particular with the
energy-momentum law $\nabla_\nu T^{\mu\nu}=0$ for matter. The
modified field equations in standard notation (see, e.g.,
\cite{15}) and signature $(+---)$ are
\begin{equation}
G_{\mu\nu} = 8\pi G T_{\mu\nu} + \Lambda g_{\mu\nu}.
\end{equation}
The cosmological term is, in four dimensions, the only possible
complication of the field equations if no higher than second order
derivatives of the metric are allowed (Lovelock theorem). This
remarkable uniqueness is one of the most attractive features of
general relativity. (In higher dimensions additional terms
satisfying this requirement are allowed.)

For the static Einstein universe the field equations (1) imply the
two relations
\begin{equation}
8\pi G \rho = \frac{1}{a^2} = \Lambda,
\end{equation}
where $\rho$ is the mass density of the dust filled universe (zero
pressure) and $a$ is the radius of curvature. (We remark, in
passing, that the Einstein universe is the only static dust
solution; one does not have to assume isotropy or homogeneity. Its
instability was demonstrated by Lema\^{\i}tre in 1927.) Einstein
was very pleased by this direct connection between the mass
density and geometry, because he thought that this was in accord
with Mach's philosophy. (His enthusiasm for what he called Mach's
principle later decreased. In a letter to F.Pirani he wrote in
1954: {\it ``As a matter of fact, one should no longer speak of
Mach's principle at all. (Von dem Machschen Prinzip sollte man
eigentlich \"uberhaupt nicht mehr sprechen''.)} \cite{4})

Einstein concludes with the following sentences:
\begin{quote}
{\it ``In order to arrive at this consistent view, we admittedly
had to introduce an extension of the field equations of
gravitation which is not justified by our actual knowledge of
gravitation. It has to be emphasized, however, that a positive
curvature of space is given by our results, even if the
supplementary term is not introduced. That term is necessary only
for the purpose of making possible a quasi-static distribution of
matter, as required by the fact of the small velocities of the
stars.''}
\end{quote}

\section{From static to expanding world models}

In the same year, 1917, de Sitter discovered a completely
different static cosmological model which also incorporated the
cosmological constant, but was {\it anti-Machian}, because it
contained no matter \cite{5}. The model had one very interesting
property: For light sources moving along static world lines there
is a gravitational redshift, which became known as the {\it de
Sitter effect}. This was thought to have some bearing on the
redshift results obtained by Slipher. Because the fundamental
(static) worldlines in this model are not geodesic, a freely-
falling particle released by any static observer will be seen by
him to accelerate away, generating also local velocity (Doppler)
redshifts corresponding to {\it peculiar velocities}. In the
second edition of his book \cite{6}, published in 1924, Eddington
writes about this:
\begin{quote}
{\it ``de Sitter's theory gives a double explanation for this
motion of recession; first there is a general tendency to scatter
(...); second there is a general displacement of spectral lines to
the red in distant objects owing to the slowing down of atomic
vibrations (...), which would erroneously be interpreted as a
motion of recession.''}
\end{quote}
I do not want to enter into all the confusion over the de Sitter
universe. This has been described in detail elsewhere (see, e.g.,
\cite{7}). An important discussion of the redshift of galaxies in
de Sitter's model by H. Weyl \cite{8} in 1923 should, however, be
mentioned. Weyl introduced an expanding version of the de Sitter
model\footnote{I recall that the de Sitter model has many
different interpretations, depending on the class of fundamental
observers that is singled out.}. For {\it small} distances his
result reduced to what later became known as the Hubble law.

Until about 1930 almost everybody {\it knew} that the Universe was
static, in spite of the two fundamental papers by Friedmann
\cite{9} in 1922 and 1924 and Lema\^{\i}tre's independent work
\cite{10} in 1927. These path breaking papers were in fact largely
ignored. The history of this early period has -- as is often the
case -- been distorted by some widely read documents. Einstein too
accepted the idea of an expanding Universe only much later. After
the first paper of Friedmann, he published a brief note claiming
an error in Friedmann's work; when it was pointed out to him that
it was his error, Einstein published a retraction of his comment,
with a sentence that luckily was deleted before publication: {\it
``[Friedmann's paper] while mathematically correct is of no
physical significance''}. In comments to Lema\^{\i}tre during the
Solvay meeting in 1927, Einstein again rejected the expanding
universe solutions as physically unacceptable. According to
Lema\^{\i}tre, Einstein was telling him: {\it ``Vos calculs sont
corrects, mais votre physique est abominable''}. On the other
hand, I found in the archive of the ETH many years ago a postcard
of Einstein to Weyl from 1923 with the following interesting
sentence: {\it ``If there is no quasi-static world, then away with
the cosmological term''}. This shows once more that history is not
as simple as it is often presented.

It also is not well-known that Hubble interpreted his famous
results on the redshift of the radiation emitted by distant
`nebulae' in the framework of the de Sitter model. He wrote:
\begin{quote}
{\it ``The outstanding feature however is that the
velocity-distance relation may represent the de Sitter effect and
hence that numerical data may be introduced into the discussion of
the general curvature of space. In the de Sitter cosmology,
displacements of the spectra arise from two sources, an apparent
slowing down of atomic vibrations and a tendency to scatter. The
latter involves a separation and hence introduces the element of
time. The relative importance of the two effects should determine
the form of the relation between distances and observed
velocities.''}
\end{quote}
 However, Lema\^{\i}tre's successful explanation of Hubble's
discovery finally changed the viewpoint of the majority of workers
in the field. At this point Einstein rejected the cosmological
term as superfluous and no longer justified  \cite{11}. He
published his new view in the {\it Sitzungsberichte der
Preussischen Akademie der Wissenschaften.} The correct citation
is:

 \begin{center}
 Einstein. A. (1931). Sitzungsber. Preuss. Akad. Wiss. 235-37.
 \end{center}

Many authors have quoted this paper but never read it. As a
result, the quotations gradually changed in an interesting, quite
systematic fashion. Some steps are shown in the following
sequence:
 \begin{enumerate}
 \item[-]{A. Einstein. 1931. Sitzsber. Preuss. Akad. Wiss. ...}
 \item[-]{A. Einstein. Sitzber. Preuss. Akad. Wiss. ... (1931)}
 \item[-]{A. Einstein (1931). Sber. preuss. Akad. Wiss. ...}
 \item[-]{Einstein. A .. 1931. Sb. Preuss. Akad. Wiss. ...}
 \item[-]{A. Einstein. S.-B. Preuss. Akad. Wis. ...1931}
 \item[-]{A. Einstein. S.B. Preuss. Akad. Wiss. (1931) ...}
 \item[-]{Einstein, A., and Preuss, S.B. (1931). Akad. Wiss. \textbf{235}}
 \end{enumerate}

Presumably, one day some historian of science will try to find out
what happened with the young physicist S.B. Preuss, who apparently
wrote just one important paper and then disappeared from the
scene.

At the end of the paper Einstein adds some remarks about the age
problem which was quite severe without the $\Lambda$-term, since
Hubble's value of the Hubble parameter was almost ten times too
large. Einstein is, however, not very worried and suggests two
ways out. First he says that the matter distribution is in reality
inhomogeneous and that the approximate treatment may be
illusionary. Then he adds that in astronomy one should be cautious
with large extrapolations in time.

Einstein repeated his new standpoint much later  \cite{12}, and
this was also adopted by many other influential workers, e.g., by
Pauli  \cite{13}. Whether Einstein really considered the
introduction of the $\Lambda$-term as ``the biggest blunder of his
life'' appears doubtful to me. In his published work and letters I
never found such a strong statement. Einstein discarded the
cosmological term just for simplicity reasons. For a minority of
cosmologists (O.Heckmann, for example  \cite{14}), this was not
sufficient reason. Paraphrasing Rabi, one might ask: `who ordered
it away'?

After the $\Lambda$-force was rejected by its inventor, other
cosmologists, like Eddington, retained it. One major reason was
that it solved the problem of the age of the Universe when the
Hubble time scale was thought to be only 2 billion years
(corresponding to the value $H_0 \sim 500\ km\ s^{-1} Mpc^{-1}$ of
the Hubble constant). This was even shorter than the age of the
Earth. In addition, Eddington and others overestimated the age of
stars and stellar systems.

For this reason, the $\Lambda$-term was employed again and a model
was revived which Lema\^{\i}tre had singled out from the many
solutions of the Friedmann-Lema\^{\i}tre equations\footnote{I
recall that Friedmann included the $\Lambda$-term in his basic
equations. I find it remarkable that for the negatively curved
solutions he pointed out that these may be open or compact (but
not simply connected).}. This so-called Lema\^{\i}tre hesitation
universe is closed and has a repulsive $\Lambda$-force
($\Lambda>0$), which is slightly greater than the value chosen by
Einstein. It begins with a big bang and has the following two
stages of expansion. In the first the $\Lambda$-force is not
important, the expansion is decelerated due to gravity and slowly
approaches the radius of the Einstein universe. At about the same
time, the repulsion becomes stronger than gravity and a second
stage of expansion begins which eventually inflates into a
whimper. In this way a positive $\Lambda$ was employed to
reconcile the expansion of the Universe with the age of stars.

The {\it repulsive} effect of a positive cosmological constant can
be seen from the following consequence of Einstein's field
equations for the time-dependent scale factor $a(t)$:
\begin{equation}
\ddot{a} = -\frac{4\pi G}{3}(\rho + 3p)a + \frac{\Lambda}{3}a,
\end{equation}
where $p$ is the pressure of all forms of matter.

Historically, the Newtonian analog of the cosmological term was
regarded by Einstein, Weyl, Pauli, and others as a {\it{Yukawa
term}}. This is not correct, as I now show.

For a better understanding of the action of the $\Lambda$-term it
may be helpful to consider a general static spacetime with the
metric (in adapted coordinates)
\begin{equation}
ds^2 = \varphi^2 dt^2 + g_{ik}dx^i dx^k,
\end{equation}
where $\varphi$ and $g_{ik}$ depend only on the spatial coordinate
$x^i$. The component $R_{00}$ of the Ricci tensor is given by
$R_{00} = \bar{\Delta}\varphi/ \varphi$, where $\bar{\Delta}$ is
the three-dimensional Laplace operator for the spatial metric
$-g_{ik}$ in (4) (see,e.g., \cite{15}). Let us write Eq. (1) in
the form
\begin{equation}
G_{\mu\nu} = \kappa (T_{\mu\nu} + T_{\mu\nu}^{\Lambda}) \quad\quad
(\kappa =  8\pi G),
\end{equation}
with
\begin{equation}
T_{\mu\nu}^{\Lambda} = \frac{\Lambda}{8\pi G} g_{\mu\nu}.
\end{equation}
This has the form of the energy-momentum tensor of an ideal fluid,
with energy density $\rho_\Lambda = \Lambda/8\pi G$ and pressure
$p_\Lambda = -\rho_\Lambda$. For an ideal fluid at rest Einstein's
field equation implies
\begin{equation}
\frac{1}{\varphi} \bar{\Delta} \varphi = 4 \pi G \Bigl[ (\rho +
3p) + \underbrace{(\rho_\Lambda + 3p_\Lambda)}_{-2\rho_\Lambda}
\Bigr].
 \end{equation}
Since the energy density and the pressure appear in the
combination $\rho + 3p$, we understand that a positive
$\rho_\Lambda$ leads to a repulsion (as in (3)). In the Newtonian
limit we have $\varphi \simeq 1 + \phi \; (\phi$ : Newtonian
potential) and $p\ll\rho$, hence we obtain the modified Poisson
equation
\begin{equation}
\Delta\phi = 4\pi G(\rho - 2\rho_\Lambda).
\end{equation}
This is the correct Newtonian limit.

As a result of revised values of the Hubble parameter and the
development of the modern theory of stellar evolution in the
1950s, the controversy over ages was resolved and the
$\Lambda$-term became again unnecessary. (Some tension remained
for values of the Hubble parameter at the higher end of recent
determinations.)

However, in 1967 it was revived again in order to explain why
quasars appeared to have redshifts that concentrated near the
value $z=2$. The idea was that quasars were born in the hesitation
era  \cite{16}. Then quasars at greatly different distances can
have almost the same redshift, because the universe was almost
static during that period. Other arguments in favor of this
interpretation were based on the following peculiarity. When the
redshifts of emission lines in quasar spectra exceed 1.95, then
redshifts of absorption lines in the same spectra were, as a rule,
equal to 1.95. This was then quite understandable, because quasar
light would most likely have crossed intervening galaxies during
the epoch of suspended expansion, which would result in almost
identical redshifts of the absorption lines. However, with more
observational data evidence for the $\Lambda$-term dispersed for
the third time.

\section{Quantum aspects of the $\Lambda$-problem}

Let me conclude this historical review with a few remarks on the
{\it quantum aspect} of the $\Lambda$-problem. Since quantum
physicists had so many other problems, it is not astonishing that
in the early years they did not worry about this subject. An
exception was Pauli, who wondered in the early 1920s whether the
zero-point energy of the radiation field could be gravitationally
effective.

As background I recall that Planck had introduced the zero-point
energy with somewhat strange arguments in 1911. The physical role
of the zero-point energy was much discussed in the days of the old
Bohr-Sommerfeld quantum theory. From Charly Enz and Armin Thellung
-- Pauli's last two assistants -- I have learned that Pauli had
discussed this issue extensively with O.Stern in Hamburg. Stern
had calculated, but never published, the vapor pressure difference
between the isotopes 20 and 22 of Neon (using Debye theory). He
came to the conclusion that without zero-point energy this
difference would be large enough for easy separation of the
isotopes, which is not the case in reality. These considerations
penetrated into Pauli's lectures on statistical mechanics
\cite{17} (which I attended). The theme was taken up in an article
by Enz and Thellung  \cite{18}. This was originally written as a
birthday gift for Pauli, but because of Pauli's early death,
appeared in a memorial volume of Helv.Phys.Acta.

>From Pauli's discussions with Enz and Thellung we know that Pauli
estimated the influence of the zero-point energy of the radiation
field -- cut off at the classical electron radius -- on the radius
of the universe, and came to the conclusion that it ``could not
even reach to the moon''.

When, as a student, I heard about this, I checked Pauli's
unpublished\footnote{A trace of this is in Pauli's Handbuch
article  \cite{19} on wave mechanics in the section where he
discusses the meaning of the zero-point energy of the quantized
radiation field.} remark by doing the following little
calculation:

In units with $\hbar=c=1$ the vacuum energy density of the
radiation field is
\[     <\rho>_{vac} = \frac{8\pi}{(2\pi)^3}\int_0^{\omega_{max}}
    \frac{\omega}{2}\omega^2 d\omega
             =  \frac{1}{8\pi^2} \omega_{max}^4 , \]
with
\begin{displaymath}
\omega_{max} = \frac{2\pi}{\lambda_{max}} = \frac{2\pi
m_e}{\alpha}.
\end{displaymath}
The corresponding radius of the Einstein universe in Eq.(2) would
then be ($M_{pl}\equiv 1/\sqrt{G}$)
\[a = \frac{\alpha^2}{(2\pi)^{\frac{2}{3}}} \frac{M_{pl}}{m_e} \frac{1}{m_e}
\sim 31 km. \] This is indeed less than the distance to the moon.
(It would be more consistent to use the curvature radius of the
static de Sitter solution; the result is the same, up to the
factor $\sqrt{3/2}$.)

For decades nobody else seems to have worried about contributions
of quantum fluctuations to the cosmological constant, also
physicists learned after Dirac's hole theory that the vacuum state
in quantum field theory is not an empty medium, but has
interesting physical properties. As an important example I mention
the papers by Heisenberg and Euler \cite{20} in which they
calculated the modifications of Maxwell's equations due to the
polarization of the vacuum. Shortly afterwards, Weisskopf
\cite{21} not only simplified their calculations but also gave a
thorough discussion of the physics involved in charge
renormalization. Weisskopf related the modification of Maxwell's
Lagrangian to the change of the energy of the Dirac sea as a
function of slowly varying external electromagnetic fields.
Avoiding the old fashioned Dirac sea, this effective Lagrangian is
due to the interaction of a classical electromagnetic field with
the vacuum fluctuations of the electron positron field. After a
charge renormalization this change is finite and gives rise to
electric and magnetic polarization vectors of the vacuum. In
particular, the refraction index for light propagating
perpendicular to a static homogeneous magnetic field depends on
the polarization direction. This is the vacuum analog of the
well-known Cotton-Mouton effect in optics. As a result, an
initially linearly polarized light beam becomes elliptic. (In
spite of great efforts it has not yet been possible to observe
this effect.)

In the thirties, people like Weisskopf were, however, not
interested in gravity. As far as I know, the first who came back
to possible contributions of the vacuum energy density to the
cosmological constant was Zel'dovich. He discussed this issue in
two papers \cite{22} during the third renaissance period of the
$\Lambda$-term, but before the advent of spontaneously broken
gauge theories. The following remark by him is particularly
interesting. Even if one assumes completely ad hoc that the
zero-point contributions to the vacuum energy density are exactly
cancelled by a bare term, there still remain higher-order effects.
In particular, {\it gravitational} interactions between the
particles in the vacuum fluctuations are expected on dimensional
grounds to lead to a gravitational self-energy density of order
$G\mu^6$, where $\mu$ is some cut-off scale. Even for $\mu$ as low
as 1 GeV (for no good reason) this is about 9 orders of magnitude
larger than the observational bound.

This illustrates that there is something profound that we do not
understand at all, certainly not in quantum field theory ( so far
also not in string theory).  We are unable to calculate the vacuum
energy density in quantum field theories, like the Standard Model
of particle physics. But we can attempt to make what appear to be
reasonable order-of-magnitude estimates for the various
contributions. \textbf{All expectations are in gigantic conflict
with the facts.} Trying to arrange the cosmological constant to be
zero is unnatural in a technical sense. It is like enforcing a
particle to be massless, by fine-tuning the parameters of the
theory when there is no symmetry principle which implies a
vanishing mass. The vacuum energy density is unprotected from
large quantum corrections. This problem is particularly severe in
field theories with spontaneous symmetry breaking. In such models
there are usually several possible vacuum states with different
energy densities. Furthermore, the energy density is determined by
what is called the effective potential, and this is {\it
dynamically} determined. Nobody can see any reason why the vacuum
of the Standard Model we ended up as the Universe cooled, has --
for particle physics standards --  an almost vanishing energy
density. Most probably, we will only have a satisfactory answer
once we shall have a theory which successfully combines the
concepts and laws of general relativity about gravity and
spacetime structure with those of quantum theory.

For more on this, see e.g. \cite{23} (and references therein), as
well as other contributions to this meeting.

\end{document}